\documentclass[conference]{IEEEtran}
\IEEEoverridecommandlockouts

\usepackage{cite}
\usepackage{amsmath,amssymb,amsfonts}
\usepackage{algorithmic}
\usepackage{graphicx}
\usepackage{textcomp}
\usepackage{xcolor}
\usepackage{subcaption}
\usepackage{hyperref}
\usepackage{eso-pic}
\def\BibTeX{{\rm B\kern-.05em{\sc i\kern-.025em b}\kern-.08em
    T\kern-.1667em\lower.7ex\hbox{E}\kern-.125emX}}

\begin{document}

\title{Gradient-Based Defensive Poisoning Against System Identification Attacks on \\ Cyber-Physical Systems
\thanks{This work was partially funded by national funds through FCT – Fundação para a Ciência e a Tecnologia, I.P., under the LASIGE Research Unit, ref. UID/00408/2025, DOI https://doi.org/10.54499/UID/00408/2025, which includes the LASIGE Seed project ref. LASIGE-SP-2025.01. Additional support was provided by Clavis FINEP/PlatCiber, under FINEP reference 1488/22, CAPES (Finance Code 001), FAPERJ under grants E-26/204.268/2024 and E-26/260.168/2026; CNPq under grants 424622/2021-1, 403601/2023-1, 315106/2023-9, 444956/2024-7, and BPP/PQ-Sr 313335/2022-2. \\ \textsuperscript{*}Corresponding author: aodsa@ciencias.ulisboa.pt}
}

\author{
\IEEEauthorblockN{
Gonçalo Fernandes$^{1}$, 
José Cecílio$^{1}$, Victor Mattos$^{2}$, Daniel S. Menasch\'e$^{3}$, Amit Bhaya$^{2}$,
Alan Oliveira de Sá$^{1*}$}
\IEEEauthorblockA{
$^{1}$LASIGE, Faculdade de Ciências, 
Universidade de Lisboa,  
Lisbon, Portugal. \\\
$^{2}$PEE/COPPE, UFRJ, Rio de Janeiro, Brazil\\
$^{3}$Instituto de Computa\c{c}\~ao, UFRJ, Rio de Janeiro, Brazil 
}}

\AddToShipoutPictureFG*{%
  \AtPageLowerLeft{%
    \put(\LenToUnit{0.7in},\LenToUnit{0.55in}){%
      \parbox[b]{7.0in}{%
        \fontsize{7}{8}\selectfont
        \textcopyright{} 2026 IEEE.
        Personal use of this material is permitted.
        Permission from IEEE must be obtained for all
        other uses, in any current or future media,
        including reprinting/republishing this material
        for advertising or promotional purposes,
        creating new collective works, for resale or
        redistribution to servers or lists, or reuse
        of any copyrighted component of this work
        in other works.
      }%
    }%
  }%
}

\maketitle

\begin{abstract}
AI-driven attacks learn the models of cyber-physical systems (CPS) by passively observing controller and plant signals, then use this knowledge to act covertly. 
Because covertness depends on the accuracy of the identified model, degrading that accuracy is a viable countermeasure. 
This work applies Adversarial Machine Learning as a defense, injecting small perturbations into the measurements the attacker observes to degrade an artificial neural network (ANN)-based system identification (SI) attack. 
The perturbations are derived from input gradients of several independently trained identification models, aggregated into a generalized perturbation, and confined to the transient regimes that carry most of the dynamic information. 
A sparsity parameter further restricts them to the most influential samples, with negligible cost in effectiveness. 
On a simulated DC motor networked control system, the proposed perturbations induce larger prediction errors than stochastic poisoning of equal root-mean-square magnitude, and the advantage widens as the magnitude grows. 
This limits the defense's operating range, which is intended for decoy environments such as CPS honeynets. 
Additional results indicate that the designed poisoning signals can also degrade N4SID-based system identification, providing evidence of their transferability beyond the ANN-based models originally targeted.
\end{abstract}


\section{Introduction}\label{sec:intro}

Cyber-Physical Systems (CPS) integrate computational logic with physical processes in critical infrastructures, industrial plants, and defense systems. Unlike attacks on conventional IT systems, which target the confidentiality, integrity, or availability of data, attacks on CPS can directly affect physical behavior, causing equipment damage, safety hazards, or service disruption. A relevant class of such attacks uses system identification (SI) to learn the plant model from eavesdropped signals and then exploits that model to act covertly~\cite{de2017covert,ferrari2020model,pasetti2021artificial,de2018evaluation}. Artificial neural networks (ANNs) are particularly attractive to attackers in this setting because, unlike metaheuristic approaches, they require no prior knowledge of the model structure~\cite{pasetti2021artificial}.

In the case of covert model-based attacks, it is proven in~\cite{smith2015covert,mattos2025design} that the covertness depends on the accuracy of the identified model. 
It follows, therefore, that one way to mitigate covert attacks is to hinder the learning (i.e., SI) process of attack algorithms. 
This work investigates this possibility by applying Adversarial Machine Learning (AML) techniques as a defensive mechanism against ANN-based attacks, specifically. 
It is worth highlighting that, while much of the AML literature focuses on attacks against machine learning (ML) models, this work explores AML from a different perspective: as a defense mechanism against ML-based attacks.
The claim is that poisoning techniques could be used to reduce the accuracy of models learned by attackers, thereby reducing the covertness of their offensive~actions. 

This work reframes gradient-based data poisoning as a defense against ML-based attackers targeting ANN-based system identification in CPS. We propose a perturbation strategy tailored to closed-loop control signals that aggregates gradients across independently trained surrogate models, thereby reducing dependence on any individual model, confines perturbations to transient regimes following setpoint changes, and sparsifies them by modifying only the most influential samples. We also evaluate the proposed defense on a DC-motor networked-control benchmark and characterize the trade-off between the degradation induced in the attacker’s model and the intensity and footprint of the injected poisoning. 
Furthermore, we investigate whether the gradient-based poisoning signals, generated using ANN-based models and originally designed to target them, can also degrade the performance of other SI methods. Specifically, we evaluate their effectiveness against the N4SID algorithm to assess whether the impact of the proposed poisoning extends beyond ANN-based identification.

Note that the proposed poisoning mechanism is intended for decoy environments, such as CPS honeynets, rather than operational systems, where it could unintentionally affect the physical process. In a honeynet, the defender controls the signals generated by the decoy and could, in principle, produce arbitrary data. However, the perturbations must remain sufficiently subtle to preserve the apparent physical realism of the emulated process. Otherwise, even an adversary with only a coarse prior model of the plant could identify inconsistencies between the observed inputs and outputs and detect the deception. Accordingly, the design objective is to maximize the degradation of the attacker’s learned model while constraining the poisoned signals to remain physically plausible.

\section{Related work}\label{sec:rel_work}

\subsection{Intelligent attacks}\label{subsec:rel_work_intel_attack}
Intelligent attacks gather and learn information about a CPS to improve the attack's efficiency, effectiveness, or covertness while pursuing a malicious objective. 
These attacks eavesdrop the communication between the controller and the physical process to infer the system dynamics and subsequently generate carefully crafted inputs capable of degrading the system in a covert manner~\cite{de2017covert}. As a result, SI has become a fundamental component of many intelligent attacks, providing the adversary with a model from which attack strategies can be designed. 
Passive SI attacks~\cite{de2017covert,de2018evaluation,ferrari2020model,pasetti2021artificial} rely solely on data collected during normal operation, making them highly covert at the expense of possible longer data collection periods, particularly when the system spends significant time in steady state. Conversely, active SI attacks~\cite{phillips2019covert,de2020bio} deliberately excite the system by injecting carefully designed control inputs, accelerating the learning process while sacrificing stealth due to the introduction of abnormal system behaviour.

Once knowledge of the system has been acquired, intelligent attacks can influence the CPS through different mechanisms with objectives varying depending on the attacker's goals. Controlled Data Loss attacks selectively suppress information exchanged within the control loop, exploiting the system's dependence on timely data without directly modifying transmitted values \cite{ferrari2020model}. In contrast, Controlled Data Injection (CDI) attacks alter the exchanged information by introducing false measurements or control commands~\cite{de2017covert,mattos2025design,pasetti2021artificial}. Since CDI attacks directly manipulate system signals, they generally require more sophisticated attack strategies in order to remain covert while still achieving the desired degradation.

As seen, several works employ SI as an intermediate step towards covert process manipulation~\cite{de2018evaluation,pasetti2021artificial}, demonstrating how learned models can be exploited to degrade control performance while remaining difficult to detect~\cite{mattos2025design}. 
Together, these works demonstrate that the success of covert intelligent attacks relies on accurately learning the target dynamics. 

\subsection{Adversarial Machine Learning}\label{subsec:rel_work_AML}
Studies on AML generally aim at improving the security of machine learning (ML) algorithms under malicious manipulation. 
Unlike conventional ML application scenarios, which assumes that the training and testing data faithfully represent the underlying process, AML considers adversaries whose objective is to degrade model performance, manipulate model behaviour, or extract model information~\cite{vassilev2024adversarial}. 

AML attacks are commonly divided according to the stage of the learning pipeline they target. Evasion attacks manipulate the input presented to an already trained model during inference, inducing incorrect predictions while leaving the learned parameters unchanged~\cite{vassilev2024adversarial}. Representative examples include the Fast Gradient Sign Method (FGSM), which demonstrated the vulnerability of deep neural networks to gradient-based adversarial perturbations \cite{goodfellow2015explaining}, the Carlini and Wagner (C\&W) attack, which formulates adversarial example generation as an optimization problem \cite{carlini2017towards}, and the Projected Gradient Descent (PGD) attack, one of the strongest first-order adversarial methods \cite{madry2017towards}. In contrast, poisoning attacks target the training phase by modifying or injecting samples into the training dataset, causing the resulting model to learn an incorrect representation of the underlying process \cite{biggio2012poisoning,steinhardt2017certified}.

Among these attack classes, poisoning is particularly relevant to the present work. Rather than exploiting a model after deployment, poisoning attacks affect the model by corrupting the data used during training. Since the model parameters are learned by minimizing a loss function over the training set, carefully crafted perturbations can systematically bias the optimization trajectory, resulting in models with degraded performance. 
Recent work has shown that highly effective poisoning attacks can be generated through gradient-based optimization. In particular, \cite{wu2022small}, which inspires this work, showed that relatively small perturbations are sufficient to substantially degrade model performance. 
Here, however, this capability is aimed as a defensive mechanism, using poisoning to counter ANN-based SI attacks.

Although not typical AML cases, the literature reports other similar efforts to mitigate SI attacks that deserve discussion. In~\cite{de2018controller}, the authors propose a switched-control strategy to hinder the identification of the closed-loop model, which combines the controller and plant dynamics, under a metaheuristic-based SI attack. In contrast, this work uses an AML technique to hinder the identification of the plant model (solely, i.e., not the closed loop model) under an ANN-based SI attack. 
Moreover, unlike~\cite {campbell2026differential}, which provides differential-privacy guarantees to protect the graph shift operator from SI attacks via Gaussian input perturbations, our work uses targeted gradient-based poisoning to degrade ANN-based identification of a CPS plant’s dynamics.

\section{ANN-based SI Attack Model}\label{sec:attack_model}

As in \cite{de2017covert,pasetti2021artificial}, the attacker is assumed to have the capability to furtively intercept the plant input and the corresponding plant output 
without modifying, injecting, or interfering with the operation of the control loop. By collecting these measurements over time, the adversary constructs a dataset that is used to train a feedforward ANN model capable of approximating the plant dynamics. The ANN receives as input the current controller output $u_k$ and a set of delayed plant outputs $\Upsilon_n = \{y_{k-1}, y_{k-2},..., y_{k-n}\}$, wherein $n$ is the order of the plant model, assumed to be known by the attacker. The hidden layers employ Rectified Linear Unit (ReLU) activation functions. A single linear output neuron produces the prediction of the plant output at the current instant $k$.

This architecture was selected due to its ability to approximate dynamic systems while maintaining a relatively simple structure. By learning the relationship between past process behavior and future outputs, the model can be used as a predictor of the physical process. Consequently, any degradation in its predictive capability may compromise its use in covert model-based attacks (e.g.,~\cite{smith2015covert}).

\section{Poisoning as Countermeasure}\label{sec:solution}

The proposed solution is inspired by \cite{wu2022small}, which demonstrated that ANN-based time series predictors can be significantly degraded through the introduction of carefully crafted perturbations. Rather than relying on random noise, the authors exploit gradient information obtained from the target model in order to identify perturbations that maximize prediction error while remaining relatively small in magnitude.

The idea behind our work is to adapt this gradient-based poisoning methodology to affect ANN-based SI attacks in CPS. To avoid drawing the attacker's attention, the defender is not interested in causing large disturbances in the signals the attacker eavesdrops. Instead, the objective is to subtly manipulate the data used to train the attacker's models, rendering their predictions unreliable and useless to run covert attacks (e.g.,~\cite{smith2015covert}). This requirement is particularly relevant in scenarios where an attacker is passively monitoring a deliberately deployed CPS honeynet to collect data for SI. In such cases, the poisoning must be subtle to avoid arousing suspicion, ensuring that the attacker continues to regard the collected measurements as legitimate. The defender's ultimate goal is to maximize the degradation of the attacker's learned model while minimizing the intensity of the defensive poisoning. 

Unlike the original work, which focuses on poisoning general time series forecasting, here the poisoning is focused on the transient regimes of closed-loop control signals -- i.e., the time interval $T$ immediately after a change in the system setpoint signal. 
These transient timeframes contain the most dynamic information about the system and are therefore expected to have a greater influence on the model training.

The proposed defensive poisoning method focuses on the generation of a generalized perturbation based on already trained ANN models. 
Initially, a set of $\mathcal{N}$ models are independently trained by the defender (the system owner) using the same dataset while varying the random initialization seed. Since the system owner does not have any knowledge about the attacker's model, multiple seeds are used in order to create a perturbation that is more generalized (i.e., not specific to one possible model trained for the targeted plant).

Following the adversarial time-series perturbation methodology described in \cite{wu2022small}, the perturbation is generated by exploiting the sensitivity of the neural network predictions with respect to its inputs. 
For each trained model, all transient regimes contained in the dataset are fed into the model. The gradient $g_k^{(i)}$ with respect to the input of a trained model $i$ at a given instant $k$
is obtained according to (\ref{eq:delta}):
\begin{equation}
\label{eq:delta}
g_k^{(i)} = \nabla_{x_k} L_k^{(i)},
\end{equation}

where $x_k = [u_k,\Upsilon_n] = [u_k, y_{k-1}, \ldots, y_{k-n}]$ represents
the input vector provided to the trained model $i$, and $L_k$ corresponds to the mean squared prediction error.
These gradients indicate the direction in which the input signal should be modified in order to increase the prediction error of the identification model \cite{goodfellow2015explaining}. 

Since each output sample $y_j$ may appear in multiple
input vectors $x_k$ due to the delayed outputs $\Upsilon_n$, its gradient
contributions are accumulated across all regressors
containing that sample. Thus, for each trained model $i$,
a single gradient contribution is obtained for each
output sample as
\begin{equation}
    g_j^{(i)} =
    \sum_{\substack{k: \, y_j \in x_k}}
    \frac{\partial L_k^{(i)}}{\partial y_j}.
    \label{eq:sample_gradient}
\end{equation}
wherein $j$ identifies the output sample $y_j$ to be perturbed and $k$ identifies the regressors in which that sample appears.

Since each trained model generates different sample-wise gradient contributions $g_j^{(i)}$ due to possible model variations arising from different training trajectories (different seeds), the gradients from all models are aggregated. Therefore, for each output sample $y_j$ in a typical transient regime, a general gradient direction $\mathcal{G}_j$ is obtained by averaging the gradients $g_j^{(i)}$ from all trained models, according to~({\ref{eq:aggregated}}):

\begin{equation}
\label{eq:aggregated}
\mathcal{G}_j = \frac{1}{\mathcal{N}}\sum_{i=1}^{\mathcal{N}} g_j^{(i)}.
\end{equation}

This aggregation accounts for the influence of each output sample across multiple regressors. Moreover, it reduces dependence on any individual model instance, improving its generalization across different models identified by the system owner. The final perturbation is normalized according to the desired perturbation magnitude $\epsilon$, according to (\ref{eq:final_pert}):

\begin{equation}
\label{eq:final_pert}
\delta_j =
\epsilon
\frac{\mathcal{G}_j}{\max_{\substack{\forall j}}{|\mathcal{G}_j|}},
\end{equation}
where $|\mathcal{G}_j|$ corresponds to the absolute value of $\mathcal{G}_j$. 

This normalization bounds the peak amplitude of the perturbation by $\epsilon$ while preserving the relative magnitude of the aggregated gradient across the transient, so that the most sensitive samples receive proportionally larger perturbations.
The perturbation $\delta_j$ is computed for each $j\in T$.

After obtaining the perturbation vector
$\Delta=\{\delta_1,\delta_2,...,\delta_m\}$, where $m$ is the number of time instants in $T$, a thresholding step is applied to retain only the perturbation components with the largest absolute values. Let $\tau\in(0,1]$ be the proportion of components to be preserved, and define $l=\left\lceil \tau m \right\rceil$ as the number of components to be retained. 
If $S_\tau$ is the set of indices corresponding to the $l$ largest values of $|\delta_j|$, the thresholded perturbation vector $\widetilde{\Delta}=\{\widetilde{\delta}_1,\widetilde{\delta}_2,...,\widetilde{\delta}_m\}$ is given by (\ref{eq:thresholded_perturbation}):
\begin{equation}
\label{eq:thresholded_perturbation}
\widetilde{\delta}_j=
\begin{cases}
\delta_j, & j\in S_\tau,\\
0, & j\notin S_\tau,
\end{cases}
\qquad j=1,\ldots,m
\end{equation}
Thus, only the fraction $\tau$ of perturbation components with the greatest absolute magnitudes is retained, while all remaining components are set to zero.

This concentrates the perturbation on the most influential samples, allowing the amount of perturbation to be reduced according to the value chosen for the parameter $\tau$. The resulting thresholded perturbation vector $\widetilde{\Delta}$ is then added to the original clean CPS signals (e.g., in a CPS honeynet), generating poisoned signals.

\section{Use Case}\label{sec:use_case}

\subsection{Target system}\label{subsec:sistema}

To evaluate the proposed solution, a simulated DC motor networked control system is used as benchmark. The DC~motor and the PI controller, which regulates the motor's rotational speed, are the same as those used in~\cite{de2017covert,de2020bio}, and their transfer functions $G(z)$ and $C(z)$, respectively, are given by (\ref{eq:plant_controller}):
\begin{equation}
\label{eq:plant_controller}
\begin{aligned}
G(z) &= \frac{0.3379z+0.2793}
{z^2-1.5462z+0.5646},
&
C(z) &= \frac{0.1701z-0.1673}{z-1}.
\end{aligned}
\end{equation}

This benchmark is selected due to its wide use in the literature related to SI attacks in CPS. The sampling rate is $50~\mathrm{samples}/\mathrm{s}$, and the system is subject to a binary setpoint signal that randomly alternates between two operating levels: $0$ and $1~ \mathrm{rad}/\mathrm{s}$. The transitions between the low and high setpoints represent a common operation in which the motor is turned on and off throughout the simulation. This naturally forces the closed-loop process to repeatedly experience transient and steady-state operating conditions, providing a dataset that contains a large number of rising and falling transitions. Such transitions are particularly important for SI attacks, as they contain rich information about the process dynamics. 

\subsection{Dataset}\label{subsec:attack_dataset}

The dataset produced and used in this work consists of ten independent recordings collected by the attacker through passive eavesdropping of the simulated control loop. Each recording contains the plant input (i.e., controller output) and the corresponding plant output signals collected over a period of 60 seconds. As such, each dataset contains 3001 samples. The ten simulations used different instances of the random binary setpoint signal to ensure greater variability. 

In this work, this dataset is used to: (i) train the $\mathcal{N}$ models used for the construction of the perturbation vector $\widetilde\Delta$ described in Section~\ref{sec:solution}; (ii) have $\widetilde\Delta$ added to it, resulting in a poisoned dataset submitted to the attacker’s training process.

\section{Results}\label{sec:results}

\subsection{Non-poisoned attack model results}\label{subsec:attack_model_results}

Before evaluating the proposed poisoning strategy, it is first necessary to establish the performance of the baseline models trained on unmodified data. For this analysis, 30 models were trained. The adopted ANN architecture, obtained through a grid search, contains two hidden layers composed of 128 neurons each. 
Each transient regime obtained from the dataset and used for training lasts $T = 100$ samples, beginning at the onset of the DC motor’s transition between the on and off states. 
Across the 30 independently trained models, the mean global test MSE was $\mu = 1.6 \times 10^{-5}$, with a standard deviation across models $\sigma = 9.3 \times 10^{-6}$.

\begin{figure}[ht!]
    \centering
    \includegraphics[width=.6\linewidth]{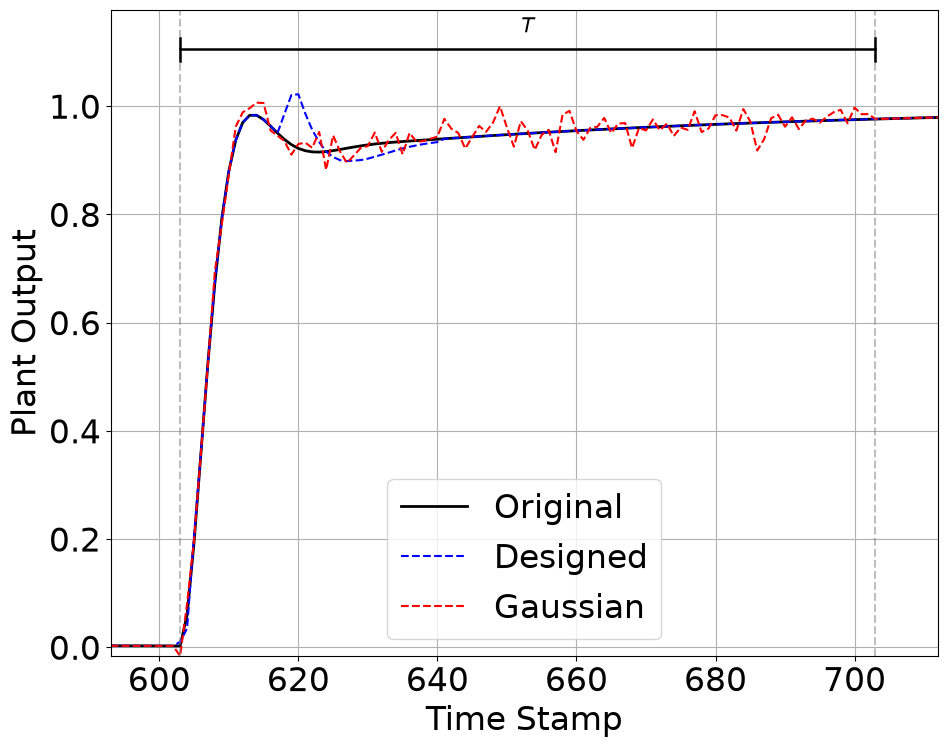}
    \caption{
    Clean and poisoned plant output over one transient window, for the proposed perturbation and for Gaussian noise of the same RMS ($\tau = 25\%$ and $\epsilon = 0.1$).
    }
    \label{fig:perturbations}
    \vspace{-5mm}
\end{figure}

\subsection{Poisoned attack model results}\label{subsec:poisoning_results}

In this section, we evaluate the effectiveness of the proposed poisoning approach by comparing the resulting model degradation with that caused by two types of noise-based poisoning signals: Gaussian noise and a pseudo-random binary sequence (PRBS). Both poisoning signals were designed to have a root mean square (RMS) magnitude equivalent to that of the proposed poisoning approach.

To compute the gradients required to construct the thresholded perturbation vector $\widetilde{\Delta}$, we used the same set of $\mathcal{N}=30$ models whose test results are reported in Section~\ref{subsec:attack_model_results}.
Their baseline performance is reported only to establish the reference error scale of an unimpeded attacker. The defender has no access to the attacker's model, which is trained separately on the poisoned data using independent seeds.
Four $\epsilon$ values were chosen to assess the impact of perturbation magnitude in our results: $0.05$, $0.10$, $0.15$, and $0.20$. 
Moreover to evaluate the influence of perturbation sparsity, three retention percentages $\tau$ are considered: $25\%$, $50\%$, and $100\%$, where the latter corresponds to $\widetilde{\Delta}=\Delta$.
As such, combining the three percentages $\tau$ with the four $\epsilon$ values yields a total of 12 perturbation profiles. 

The resulting perturbations are then applied to the original clean datasets, generating poisoned datasets. 
Fig.~\ref{fig:perturbations} illustrates the perturbation introduced by our approach using $\tau=0.25$ and $\epsilon=0.1$, alongside the Gaussian perturbation with the same~RMS.

To evaluate the impact of proposed poisoning against the ANN-based SI attacks, the created poisoned datasets are used to train a new set of 30 models with independent random seeds. 
The impact of poisoning on the inference capability of the attacker’s models is then assessed by testing the poisoned models on the original, unpoisoned dataset. This evaluation quantifies the inference error that the attacker’s poisoned models would produce when applied directly to the benchmark system. 
The degradation in identification performance is quantified by the maximum absolute error (MaxAE) observed during inference for each model over simulations of duration $T=2s$. 
The intensity of the introduced poisoning is characterized by its RMS value.

\begin{figure*}[!t]
\centering

\begin{subfigure}[t]{0.3\textwidth}
    \centering
    \includegraphics[width=\linewidth]{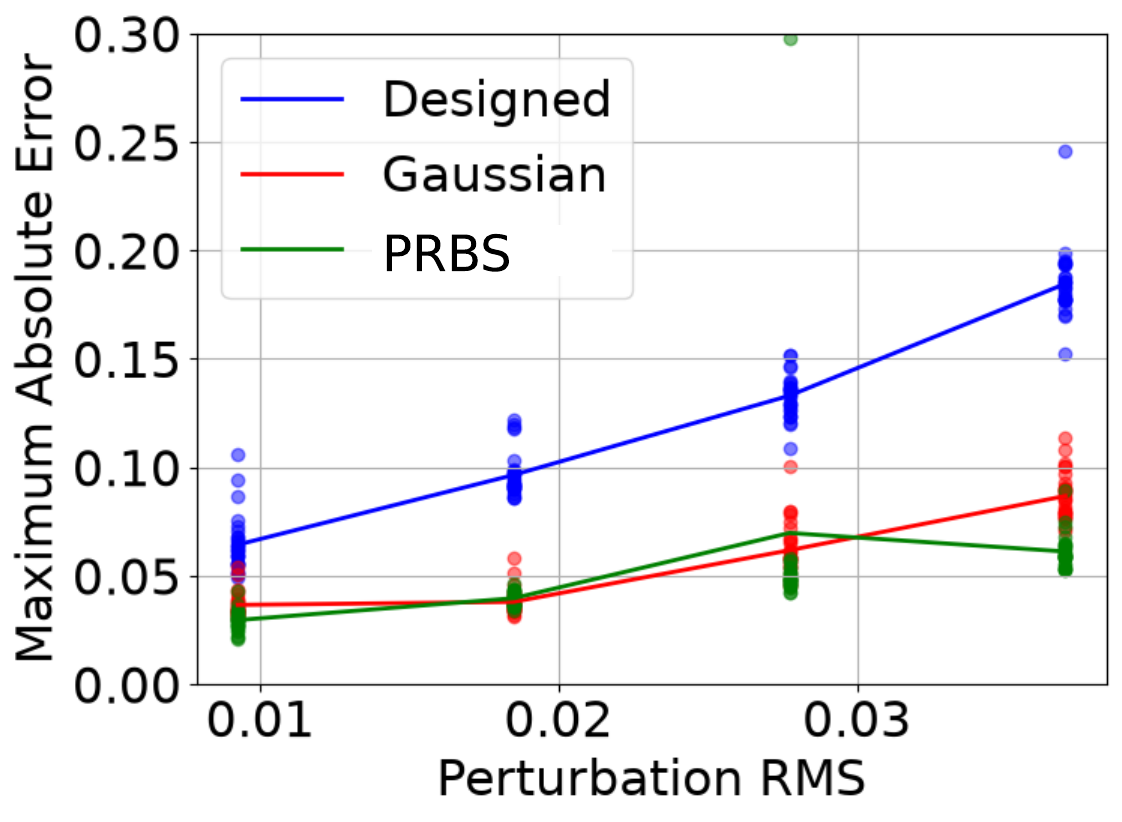}
    \caption{$\tau=25\%$}
    \label{fig:maxerror25}
\end{subfigure}
\hfill
\begin{subfigure}[t]{0.3\textwidth}
    \centering
    \includegraphics[width=\linewidth]{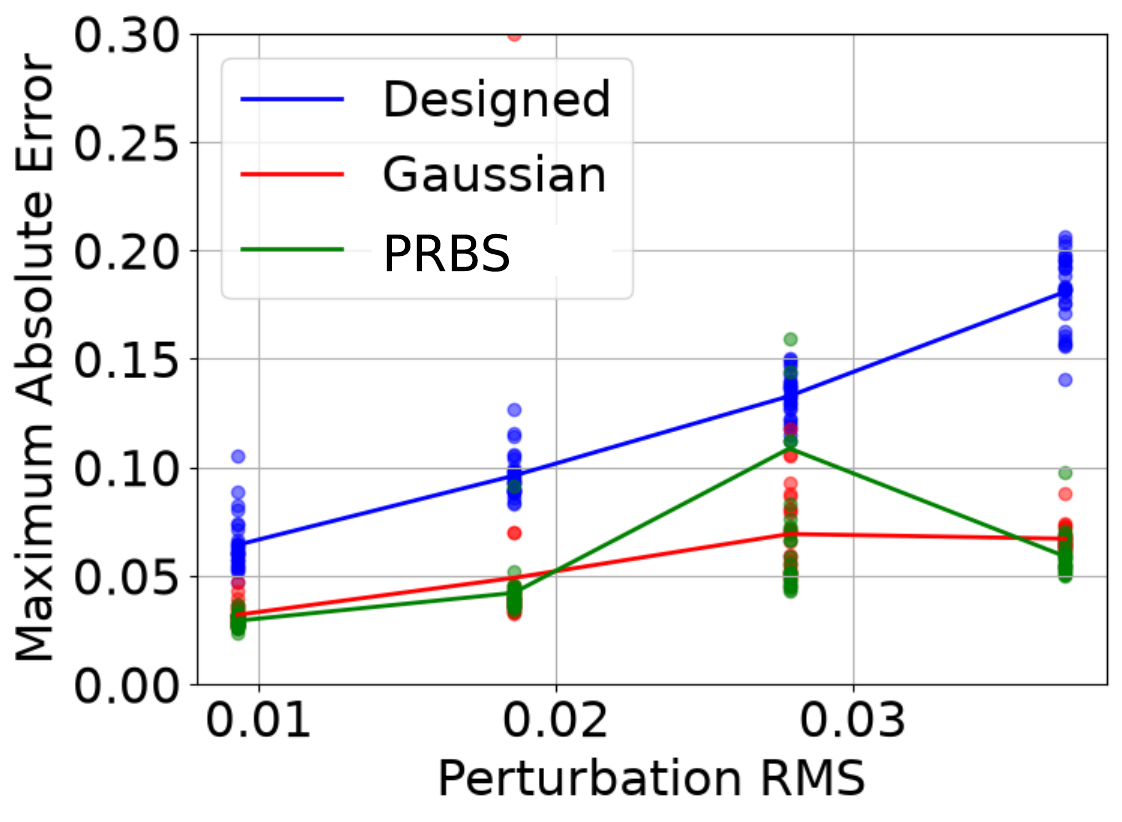}
    \caption{$\tau=50$\%}
    \label{fig:maxerror50}
\end{subfigure}
\hfill
\begin{subfigure}[t]{0.3\textwidth}
    \centering
    \includegraphics[width=\linewidth]{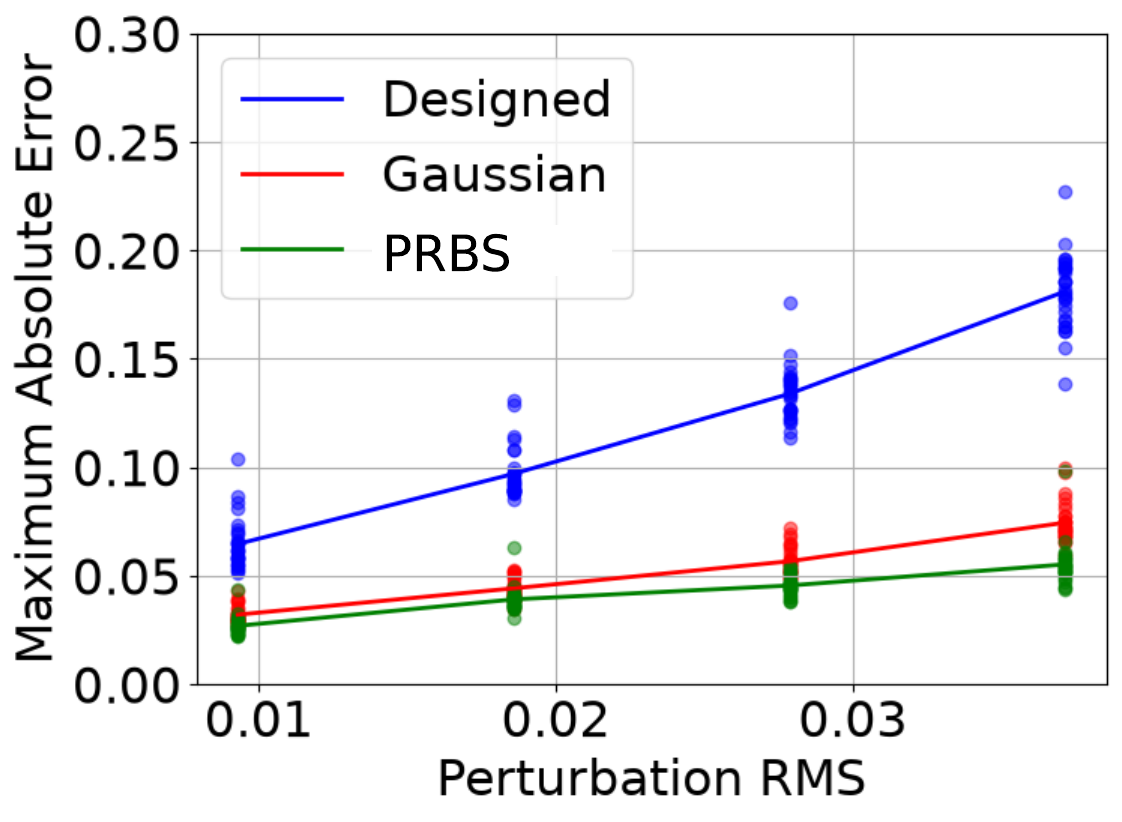}
    \caption{$\tau=100$\%}
    \label{fig:maxerror100}
\end{subfigure}

\caption{MaxAE of the poisoned ANN models for different $\tau$ and poisoning RMS values (resulting from the different $\epsilon$ values).}
\label{fig:transitionmsecomparison}
\vspace{-3mm}
\end{figure*}

\begin{figure*}[!t]
\centering

\begin{subfigure}[t]{0.3\textwidth}
    \centering
    \includegraphics[width=\linewidth]{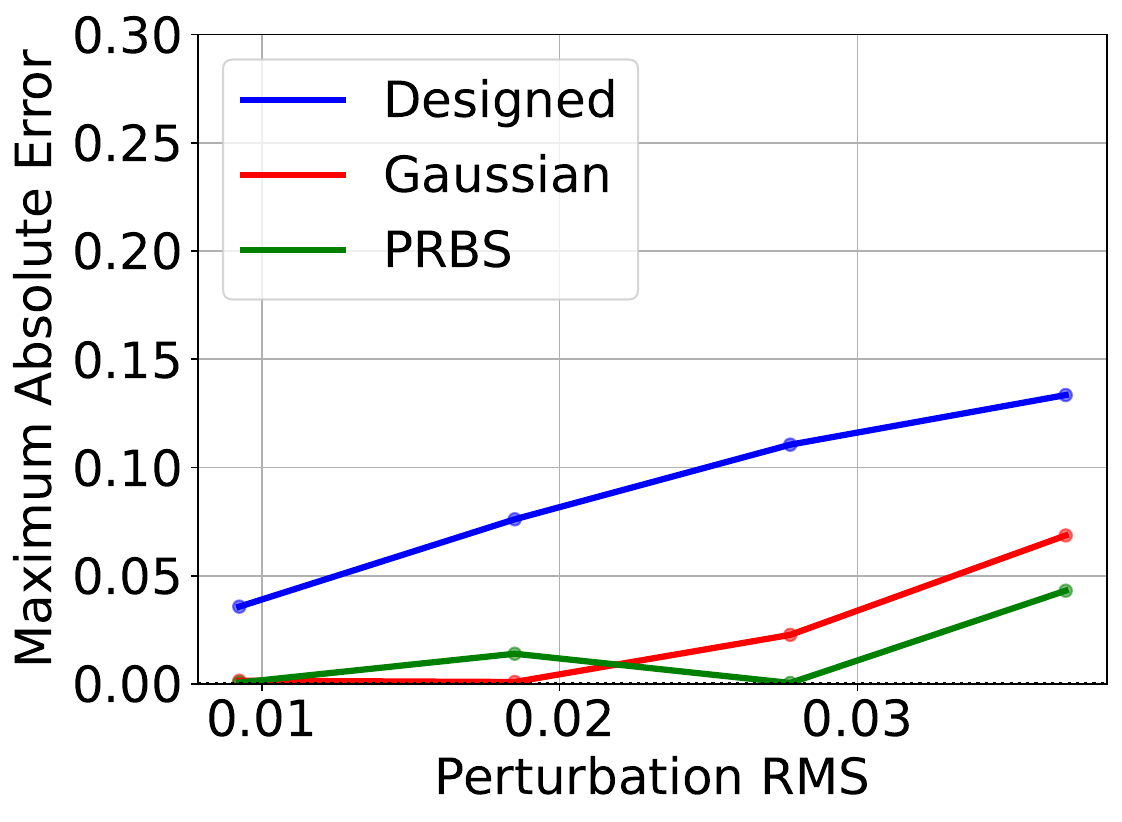}
    \caption{$\tau=25\%$}
    \label{fig:n4sid_maxerror25}
\end{subfigure}
\hfill
\begin{subfigure}[t]{0.3\textwidth}
    \centering
    \includegraphics[width=\linewidth]{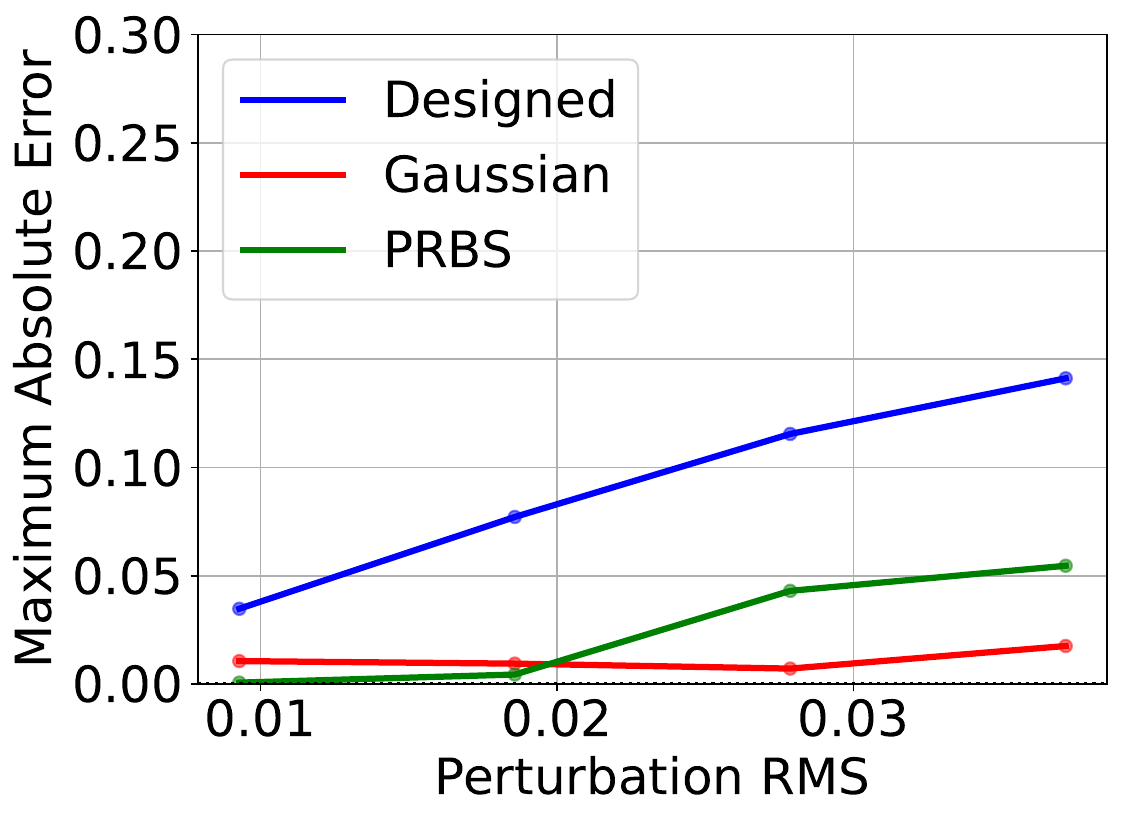}
    \caption{$\tau=50$\%}
    \label{fig:n4sid_maxerror50}
\end{subfigure}
\hfill
\begin{subfigure}[t]{0.3\textwidth}
    \centering
    \includegraphics[width=\linewidth]{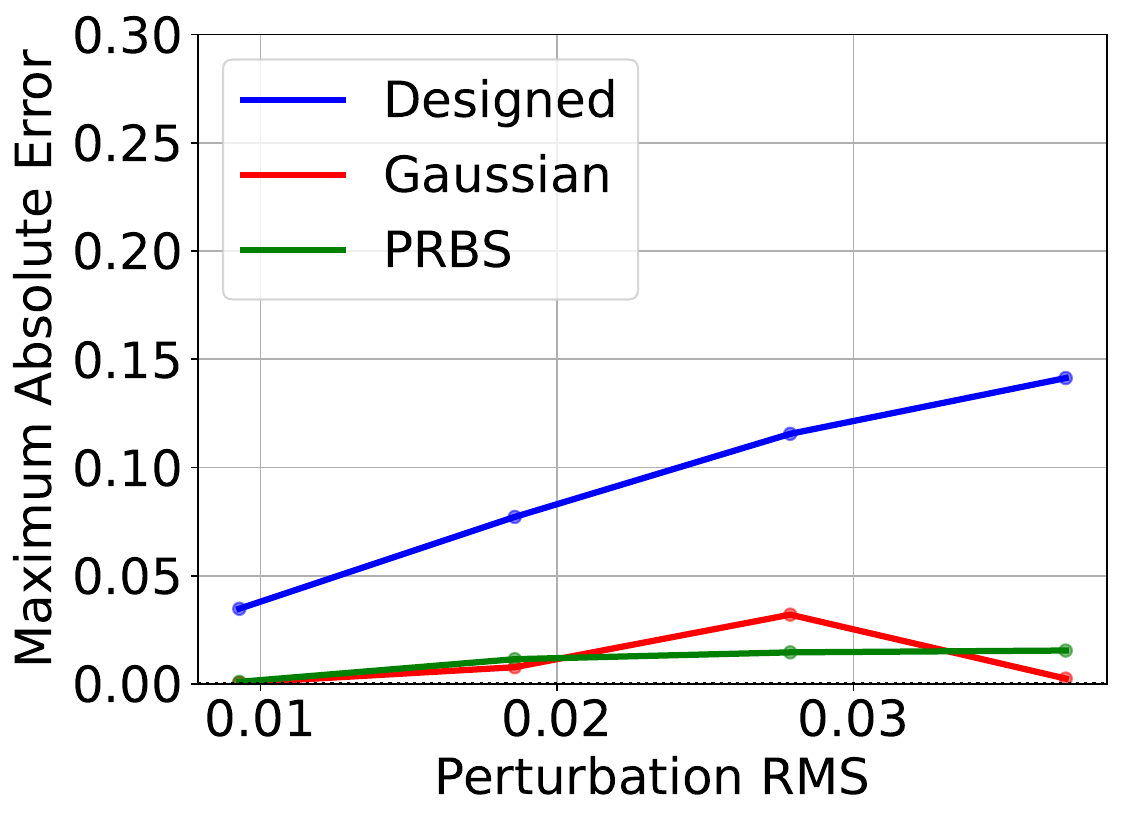}
    \caption{$\tau=100$\%}
    \label{fig:n4sid_maxerror100}
\end{subfigure}

\caption{MaxAE of the poisoned N4SID models for different $\tau$ and poisoning RMS values. 
}
\label{fig:n4sid_transfer}
\vspace{-4mm}
\end{figure*}

Fig.~\ref{fig:transitionmsecomparison} shows as dots the MaxAE reached by each retained poisoned model as a function of the perturbation RMS for the three considered retention percentages $\tau$ and the four $\epsilon$. The lines represent the mean MaxAE obtained across the retained models from 30 independent training runs. Across all $\tau$ and $\epsilon$, eight poisoned models were considered outliers, most presenting unstable behaviour: three Gaussian models and five PRBS models. These were identified as models whose global test MSE exceeded $1~(\mathrm{rad}/\mathrm{s})^2$ and were excluded from the reported results. Among the retained models, across all $\tau$, the proposed gradient-based poisoning consistently produces larger mean MaxAE than Gaussian noise and PRBS-based perturbations of equivalent RMS, with the difference generally becoming more pronounced as the perturbation magnitude $\epsilon$ increases. Within this retained subset, the proposed method is more effective at inducing large deviations in the attacker model while maintaining the same overall perturbation intensity.
Interestingly, as shown in Fig.~\ref{fig:transitionmsecomparison}, relatively sparse perturbations remain effective, even when only 25\% of the transient samples are modified. This suggests that poisoning can be concentrated on a small subset of influential samples without substantially compromising its effectiveness.

\begin{figure}
    \centering
    \includegraphics[width=0.84\linewidth]{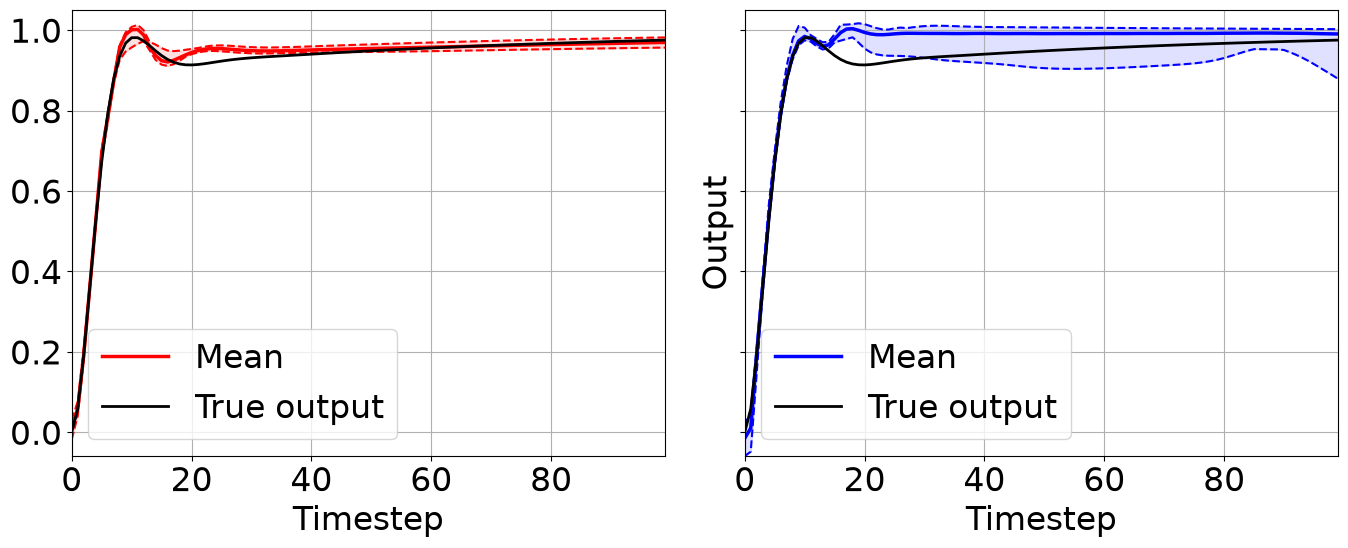}
    \caption{Prediction regions bounded by the minimum and maximum trajectories for models poisoned with Gaussian noise (red) and the proposed gradient-based poisoning (blue), with $\tau=25\%$ and $\epsilon=0.1$.}
    \label{fig:maxminmean}
    \vspace{-5mm}
\end{figure}

Fig. \ref{fig:maxminmean} further illustrates this effect by showing the prediction regions bounded by the minimum and maximum trajectories produced by models poisoned with Gaussian noise and the proposed gradient-based poisoning, with $\tau=25\%$ and $\epsilon=0.1$. It is possible to see that the models trained with the proposed poisoning exhibit a noticeably wider prediction range and larger deviations from the true output, during and particularly after the transient response. In contrast, the Gaussian-poisoned models remain more tightly concentrated around the true system trajectory. Overall, these results show that the proposed poisoning achieves a stronger degradation of the attacker's learned model than Gaussian poisoning of equivalent magnitude.\footnote{Code available at \url{https://github.com/goncalo-ap-fernandes/poisoning-repo}}

\subsection{Transferability of Poisoning to Other SI Methods}\label{subsec:results_transf}

In this section, we assess whether poisoning signals designed for ANN-based system identification also degrade the performance of a different identification method, namely N4SID~\cite{VANOVERSCHEE199475}. We apply the original perturbations without modification or adaptation and thus test whether their effectiveness transfers across identification approaches.

The plant is identified using MATLAB’s N4SID with the known system order fixed at two and all other options set to their defaults. 
For fixed data and options, N4SID requires no random initialization. Each point in Fig.~\ref{fig:n4sid_transfer} represents one fit for one supplied profile, while the ANN curves in Fig.~\ref{fig:transitionmsecomparison} average the retained models from 30 training runs. All N4SID models are stable and produce finite predictions; none is excluded by the MSE criterion.

As shown in Fig.~\ref{fig:n4sid_transfer}, the designed gradient-based perturbation yields a higher MaxAE than both RMS-matched stochastic poisoning in all 12 conditions, with the error generally increasing with RMS for each value of $\tau$. In contrast to the ANN results (Fig.\ref{fig:transitionmsecomparison} and~\ref{fig:n4sid_transfer}), the Gaussian and PRBS poisoning produce lower errors under N4SID. Under both stochastic poisonings, the poisoned ANN models show non-negligible errors even at the lowest RMS and an overall tendency to increase as the perturbation~grows.

Across the 12 conditions, the median of the MaxAE ratios between the designed perturbation and Gaussian and PRBS poisoning are $9.08$ and $8.52$, respectively, for N4SID. For the ANN models, the corresponding medians are $2.14$ and $2.42$. 
Thus, although N4SID generally has lower absolute errors, its median separation between the designed perturbation and each stochastic control is proportionally larger.

The lower N4SID errors under stochastic poisoning may partly reflect the match between its model structure and the benchmark plant, which is linear and of order two. Restricting identification to this class may limit the fitting of perturbation features that do not correspond to the plant dynamics. ANNs permit more flexible, nonlinear relationships, but also incur prediction error on clean data. The comparison therefore combines baseline identification error with the effect of poisoning and does not isolate the contribution of model~flexibility.

The designed perturbation may affect N4SID because its nearly opposite profiles during rising and falling transitions repeatedly distort the input-output relationship in a consistent way. A linear identifier could incorporate this distortion as a change in plant dynamics.

\section{Conclusion}\label{sec:concl}

The results show that the proposed defensive poisoning strategy degrades ANN-based system-identification attacks more effectively than Gaussian and PRBS-based perturbations with the same RMS, without increasing the number of altered samples. This suggests that gradient-informed perturbations can increase the attacker’s model error while preserving a limited poisoning footprint. 
Furthermore, the results provide evidence of poisoning transferability across ANN-based and N4SID system identification methods, although its effectiveness against other identification approaches remains to be investigated.
However, these findings have several limitations. The evaluation assumes that the defender’s surrogate models and the attacker’s models share the same architecture, input ordering, and training procedure. 
Moreover, stealth is assessed only through perturbation RMS and the number of modified samples. 

Because the proposed perturbations are sparse, deterministic, and aligned with setpoint transitions, they may be detectable by structure-aware analyses even when their energy matches that of white noise; in particular, averaging aligned transients could reinforce, rather than cancel, the perturbation. Finally, the evaluation measures prediction error in the poisoned model and does not directly assess the detectability or operational impact of an attack derived from that model. 
Future work should therefore further investigate poisoning transferability across a broader range of SI methods and systems, randomization across transients, additional signal types, and direct measures of attack detectability.

\bibliographystyle{IEEEtran}
\bibliography{references}

@article{smith2015covert,
  title={Covert misappropriation of networked control systems: Presenting a feedback structure},
  author={Smith, Roy S},
  journal={IEEE Control Systems Magazine},
  volume={35},
  number={1},
  pages={82--92},
  year={2015},
  publisher={IEEE}
}

@article{wu2022small,
  title={Small perturbations are enough: Adversarial attacks on time series prediction},
  author={Wu, Tao and Wang, Xuechun and Qiao, Shaojie and Xian, Xingping and Liu, Yanbing and Zhang, Liang},
  journal={Information Sciences},
  volume={587},
  pages={794--812},
  year={2022},
  publisher={Elsevier}
}

@article{de2017covert,
  title={Covert attacks in cyber-physical control systems},
  author={de S{\'a}, Alan Oliveira and da Costa Carmo, Luiz F Rust and Machado, Raphael CS},
  journal={IEEE Transactions on Industrial Informatics},
  volume={13},
  number={4},
  pages={1641--1651},
  year={2017},
  publisher={IEEE}
}

@inproceedings{de2018evaluation,
  title={Evaluation on passive system identification and covert misappropriation attacks in large pressurized heavy water reactors},
  author={de S{\'a}, Alan Oliveira and Carmo, Luiz Fernando Rust da C and Machado, Raphael C Santos},
  booktitle={2018 Workshop on Metrology for Industry 4.0 and IoT},
  pages={203--208},
  year={2018},
  organization={IEEE}
}

@article{pasetti2021artificial,
  title={Artificial neural network-based stealth attack on battery energy storage systems},
  author={Pasetti, Marco and Ferrari, Paolo and Bellagente, Paolo and Sisinni, Emiliano and de S{\'a}, Alan Oliveira and do Prado, Charles B and David, Rodrigo P and Machado, Raphael Carlos Santos},
  journal={IEEE Transactions on Smart Grid},
  volume={12},
  number={6},
  pages={5310--5321},
  year={2021},
  publisher={IEEE}
}

@article{ferrari2020model,
  title={Model-based stealth attack to networked control system based on real-time Ethernet},
  author={Ferrari, Paolo and Sisinni, Emiliano and Bellagente, Paolo and Rinaldi, Stefano and Pasetti, Marco and De Sa, Alan Oliveira and Machado, Raphael and Carmo, Luiz and Casimiro, Ant{\'o}nio},
  journal={IEEE Trans. on Industrial Electronics},
  volume={68},
  number={8},
  pages={7672--7683},
  year={2021},
  publisher={IEEE}
}

@inproceedings{phillips2019covert,
  title={A covert system identification attack on constant setpoint control systems},
  author={Phillips, Tyler and Mehrpouyan, Hoda and Gardner, John and Reese, Stephen},
  booktitle={2019 Seventh International Symposium on Computing and Networking Workshops (CANDARW)},
  pages={367--373},
  year={2019},
  organization={IEEE}
}

@article{de2020bio,
  title={Bio-inspired active system identification: a cyber-physical intelligence attack in networked control systems},
  author={de S{\'a}, Alan Oliveira and Carmo, Luiz and Machado, Raphael},
  journal={Mobile Networks and Applications},
  volume={25},
  number={5},
  pages={1944--1957},
  year={2020},
  publisher={Springer}
}

@article{steinhardt2017certified,
  title={Certified defenses for data poisoning attacks},
  author={Steinhardt, Jacob and Koh, Pang Wei W and Liang, Percy S},
  journal={Advances in neural information processing systems},
  volume={30},
  year={2017}
}

@inproceedings{mattos2025design,
  title={Design and Detection of Covert Man-in-the-Middle Cyberattacks on Water Treatment Plants},
  author={Mattos, Victor and Schmidt, Jo{\~a}o and Bhaya, Amit and de S{\'a}, Alan Oliveira and Menasche, Daniel and Srivastava, Gaurav},
  booktitle={Proceedings of the 2025 Workshop on Re-design Industrial Control Systems with Security},
  pages={7--15},
  year={2025}
}

@techreport{vassilev2024adversarial,
  author      = {Vassilev, Apostol and Oprea, Alina and Fordyce, Alie and Anderson, Hyrum},
  title       = {Adversarial Machine Learning: A Taxonomy and Terminology of Attacks and Mitigations},
  institution = {NIST},
  type        = {Report},
  number      = {NIST AI 100-2e2023},
  address     = {Gaithersburg, MD},
  year        = {2024},
  month       = jan,
}

@inproceedings{goodfellow2015explaining,
  title     = {Explaining and Harnessing Adversarial Examples},
  author    = {Goodfellow, Ian J. and Shlens, Jonathon and Szegedy, Christian},
  booktitle = {International Conference on Learning Representations},
  year      = {2015},
  eprint    = {1412.6572},
  archivePrefix = {arXiv},
  primaryClass  = {stat.ML}
}

@INPROCEEDINGS{carlini2017towards,
  author={Carlini, Nicholas and Wagner, David},
  booktitle={2017 IEEE Symposium on Security and Privacy (SP)}, 
  title={Towards Evaluating the Robustness of Neural Networks}, 
  year={2017},
  volume={},
  number={},
  pages={39-57},
  doi={10.1109/SP.2017.49}}

@inproceedings{madry2017towards,
  title={Towards Deep Learning Models Resistant to Adversarial Attacks},
  author={Madry, Aleksander and Makelov, Aleksandar and Schmidt, Ludwig and Tsipras, Dimitris and Vladu, Adrian},
  booktitle={International Conference on Learning Representations},
  year={2018}
}

@inproceedings{biggio2012poisoning,
  author    = {Biggio, Battista and Nelson, Blaine and Laskov, Pavel},
  title     = {Poisoning Attacks against Support Vector Machines},
  booktitle = {Proceedings of the 29th Int. Conference on Machine Learning},
  pages     = {1467--1474},
  year      = {2012},
  address   = {Edinburgh, Scotland, UK},
  publisher = {Omnipress},
  doi       = {10.5555/3042573.3042761},
  eprint    = {1206.6389},
  archivePrefix = {arXiv},
  primaryClass  = {cs.LG}
}

@article{de2018controller,
  title={A controller design for mitigation of passive system identification attacks in networked control systems},
  author={de S{\'a}, Alan O and da Costa Carmo, Luiz FR and Machado, Raphael CS},
  journal={Journal of Internet Services and Applications},
  volume={9},
  number={1},
  pages={2},
  year={2018},
  publisher={Springer}
}

@inproceedings{campbell2026differential,
  title={Differential Privacy of Network Parameters From a System Identification Perspective},
  author={Campbell, Andrew and Scaglione, Anna and Liu, Hang and Elvira, V{\'\i}ctor and Peisert, Sean and Arnold, Daniel},
  booktitle={ICASSP 2026-2026 IEEE International Conference on Acoustics, Speech and Signal Processing},
  pages={441--445},
  year={2026},
  organization={IEEE}
}

@article{VANOVERSCHEE199475,
title = {N4SID: Subspace algorithms for the identification of combined deterministic-stochastic systems},
journal = {Automatica},
volume = {30},
number = {1},
pages = {75-93},
year = {1994},
issn = {0005-1098},
doi = {10.1016/0005-1098(94)90230-5},
author = {Peter {Van Overschee} and Bart {De Moor}},
}
\end{document}